\documentclass[manuscript]{emulateapj}
\usepackage{times}
\usepackage{amsmath}
\usepackage[usenames,dvipsnames]{color}
\usepackage{multirow}
\usepackage{soul}
\usepackage{comment}

\newcommand{\myemail}{mathieu.powalka@astro.unistra.fr}

\slugcomment{To appear in Astrophysical Journal}

\shorttitle{NGVS. XXXII: Indication for a Globular Cluster Substructure in the Virgo Galaxy Cluster Core}
\shortauthors{Powalka et al.}

\begin{document}

\title{Next Generation Virgo Cluster Survey (NGVS). XXXII:~Search for a Globular Cluster Substructure in the Virgo Galaxy Cluster Core}

\author{Mathieu Powalka$^{1}$, Thomas H.~Puzia$^{2}$, Ariane~Lan\c{c}on$^{1}$, Alessia Longobardi$^{3}$
Eric W. Peng$^{3,4}$, Pierre-Alain Duc$^{1}$,
Karla Alamo-Mart\'inez$^{2}$, John P. Blakeslee$^{5}$, Patrick C\^{o}t\'e$^{5}$, Jean-Charles Cuillandre$^{6}$, Patrick Durrell$^{7}$, Paul Eigenthaler$^{2}$, Laura Ferrarese$^{5}$, Puragra Guhathakurta$^{8}$, S. D. J. Gwyn$^{5}$, Patrick Hudelot$^{9}$ Chengze Liu$^{10,11}$, Simona Mei$^{12,13,14}$, Roberto~P.~Mu\~noz$^{2}$, Joel Roediger$^{5}$, Rub\'en S\'anchez-Janssen$^{15}$, Elisa Toloba$^{16}$, Hongxin Zhang$^{2}$
}
\email{\myemail}
\altaffiltext{1}{Observatoire Astronomique de Strasbourg, Universit\'e de Strasbourg, CNRS, UMR 7550, 11 rue de l'Universit\'e, F-67000 Strasbourg, France}
\altaffiltext{2}{Institute of Astrophysics, Pontificia Universidad Cat\'olica de Chile, Av.~Vicu\~na Mackenna 4860, 7820436 Macul, Santiago, Chile}
 \altaffiltext{3}{Kavli Institute for Astronomy and Astrophysics, Peking University, Beijing 100871, China}
 \altaffiltext{4}{Department of Astronomy, Peking University, Beijing 100871, China}
 \altaffiltext{5}{National Research Council of Canada, Herzberg Astronomy and Astrophysics Research Centre, 5071 West Saanich Road, Victoria, BC, V9E 2E7, Canada}
 \altaffiltext{6}{AIM Paris Saclay, CNRS/INSU, CEA/Irfu, Universit\'e Paris Diderot, Orme des Merisiers, F-91191 Gif-sur-Yvette Cedex, France}
 \altaffiltext{7}{Department of Physics and Astronomy, Youngstown State University, One University Plaza, Youngstown, OH 44555, USA}
 \altaffiltext{8}{UCO/Lick Observatory, Department of Astronomy and Astrophysics, University of California Santa Cruz, 1156 High Street, Santa Cruz, CA 95064, USA}
 \altaffiltext{9}{Institut d'Astrophysique de Paris, UMR 7095 CNRS \& UPMC, 98bis Bd Arago, F-75014 Paris, France}
 \altaffiltext{10}{Center for Astronomy and Astrophysics, Department of Physics and Astronomy, Shanghai Jiao Tong University, Shanghai 200240, China}
\altaffiltext{11}{Shanghai Key Lab for Particle Physics and Cosmology, Shanghai Jiao Tong University, Shanghai 200240, China}
\altaffiltext{12}{LERMA, Observatoire de Paris, PSL Research University, CNRS, Sorbonne Universit\'es, UPMC Univ. Paris 06, F-75014 Paris, France}
\altaffiltext{13}{Universit\'{e} Paris Denis Diderot, Universit\'e Paris Sorbonne Cit\'e, 75205 Paris Cedex13, France}
\altaffiltext{14}{Jet Propulsion Laboratory, Cahill Center for Astronomy \& Astrophysics, California Institute of Technology, 4800 Oak Grove Drive, Pasadena, California, USA}
\altaffiltext{15}{UK Astronomy Technology Centre, Royal Observatory Edinburgh, Blackford Hill, Edinburgh, EH9 3HJ, UK}
\altaffiltext{16}{Department of Physics, University of the Pacific, 3601 Pacific Avenue, Stockton, CA 95211, USA}

\begin{abstract}
Substructure in globular cluster (GC) populations around large galaxies is expected in galaxy formation scenarios that involve accretion or merger events, and it has been searched for using direct associations between GCs and structure in the diffuse galaxy light, or with GC kinematics. Here, we present a search for candidate substructures in the GC population around the Virgo cD galaxy M87 through the analysis of the spatial distribution of the GC colors.~The study is based on a sample of $\sim\!1800$ bright GCs with high-quality $u,g,r,i,z,K_s$ photometry, selected to ensure a low contamination by foreground stars or background galaxies.~The spectral energy distributions of the GCs are associated with formal estimates of age and metallicity, which are representative of its position in a 4-D color-space relative to standard single stellar population models.~Dividing the sample into broad bins based on the relative formal ages, we observe inhomogeneities which reveal signatures of GC substructures.~The most significant of these is a spatial overdensity of GCs with relatively young age labels, of diameter $\sim\!0.1$\,deg ($\sim\!30\,$kpc), located to the south of M87.~The significance of this detection is larger than about 5$\sigma$ after accounting for estimates of random and systematic errors.~Surprisingly, no large Virgo galaxy is present in this area, that could potentially host these GCs.~But candidate substructures in the M87 halo with equally elusive hosts have been described based on kinematic studies in the past.~The number of GC spectra available around M87 is currently insufficient to clarify the nature of the new candidate substructure.
\end{abstract}


\section{Introduction}

In order to unravel the history of a present-day galaxy, one may try to disassemble all the observable remnants of past interactions and trace back the events that occurred since the galaxy's formation.~Globular clusters (GC), which can be observed in a large volume of the local universe, are interesting tracers in this context.

Previous seminal studies suggested that GCs in massive galaxies could be formed in dense star-forming regions of galaxy mergers \citep{ashman1992} or be accreted from disrupted infalling galaxies \citep{cote1998}.~Evidence for these processes is seen both in the Milky Way \citep[e.g.][]{keller2012}, and in external galaxies \citep[e.g.][]{whitmore93,lim2017}.~Violent galaxy interactions such as major mergers are known to leave morphological signatures for as long as a few Gyr \citep[$\sim\,$2-3\,Gyr for a major merger; e.g.][]{borne1991}.~When morphological features in the diffuse light fade away, structure may still be detected in the kinematic phase space of the GC population \citep{romanowsky12}.~As the GCs keep the chemical imprints of their birth place, studies of their stellar populations also help reconstructing the history of their assembly around a host. In the Milky Way, the shape of the horizontal branch has been an early argument in favor of a dichotomy between accretion and in-situ formation \citep{SearleZinn1978}. Further out, more or less bimodal color distributions of populations of unresolved GCs point to a combination of the two scenarios. But one or two colors are insufficient to capture the whole complexity of possible stellar population properties \citep{powalka2016b}, and a more complete usage of the SED may allow us unravel more details of GC assembly histories.

In earlier articles of this series \citep[][hereafter referred to as Papers I and II]{powalka2016b, powalka2017}, we studied the GCs in the Virgo core region (around M87) using the data of the {\it Next Generation Virgo cluster Survey} \cite[NGVS/NGVS-IR; see][]{ferrarese2012, munoz2014}.~We confronted their $u^*$,$g$,$r$,$i$,$z$,$K_s$ photometry with the predictions of single stellar population (SSP) models for a broad grid of ages and metallicities.~Although the {\it absolute} ages and metallicities one may assign to the GCs are affected by biases and should be manipulated with great caution, we emphasized that the {\it relative} values of these assigned labels contain rich information on the relative position of the GCs in color-space, and may be sensitive to subtle differences such as those induced by modified abundance ratios.

Therefore, in this letter, we analyze the spectral energy distributions (SEDs) of Virgo-core GCs using derived quantities (mainly formal labels of age and metallicity) from our broadband photometric filter set.~Knowing that the cD galaxy M87 has experienced multiple mergers, we aim to search for local variations that might indicate differences in GC origins.

\begin{figure*}[t]
\begin{center}
\includegraphics[width=\textwidth]{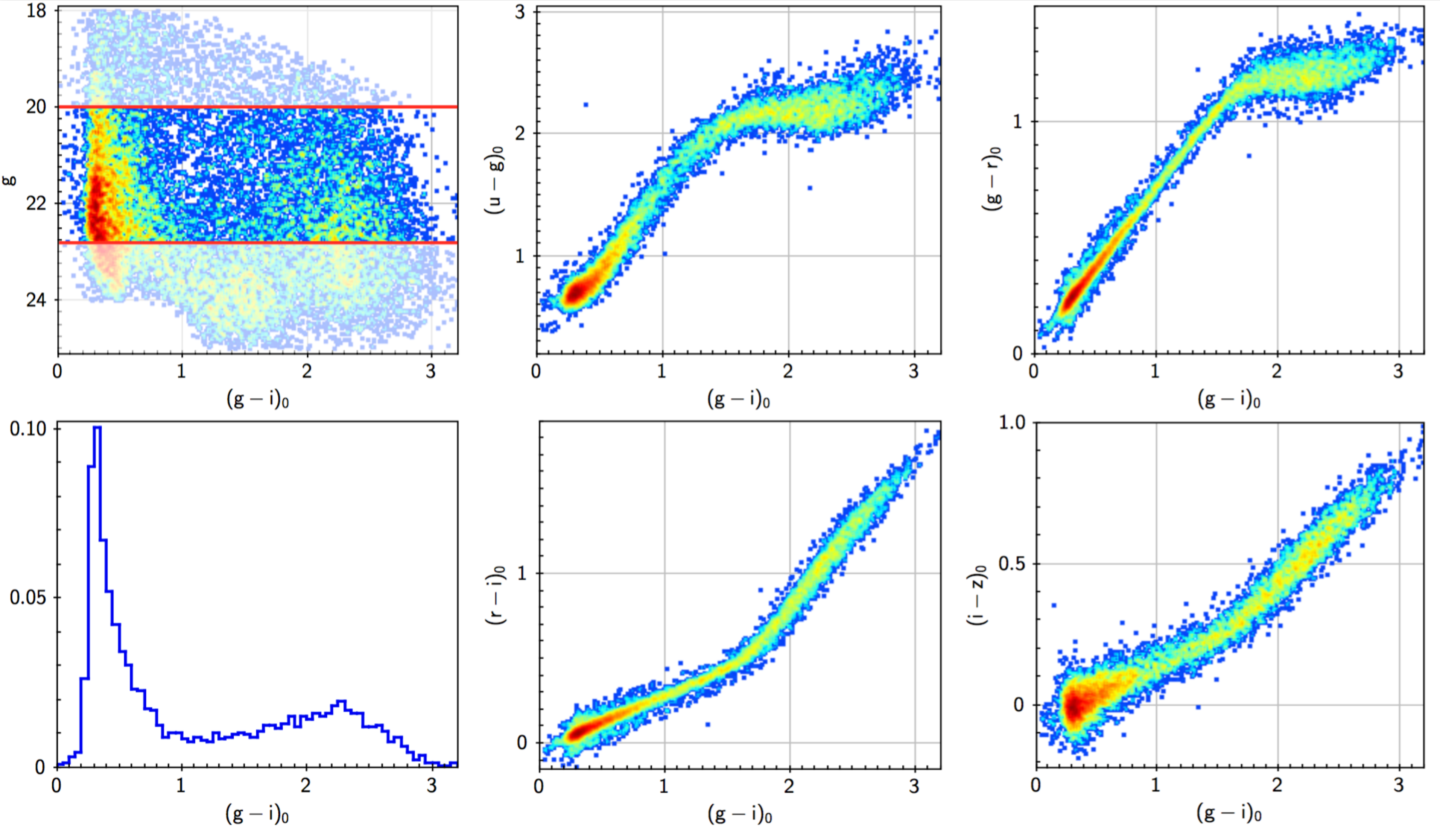}
\caption{Star selection for the study of the homogeneity of the photometry. Stars with $(g-i)_0<0.6$ are mostly turn-off stars of the Milky Way halo, and more specifically of the Virgo Overdensity and the Sagittarius stream \citep[][]{jerjen2013, durrell2014, loc16}, while red stars are mostly fainter dwarfs of the Milky Way disks. Among the stars in the first panel, we select those with $20<(g-i)_0<22.8$. The color-color diagrams and the $(g-i)_0$ histogram for the selected subset are shown in the subsequent panels (a handful of outliers seen in those diagrams are removed for the analysis).
\label{fig:starselection}}
\end{center}
\end{figure*}

\begin{figure*}[t]
\centering
\includegraphics[width=\linewidth]{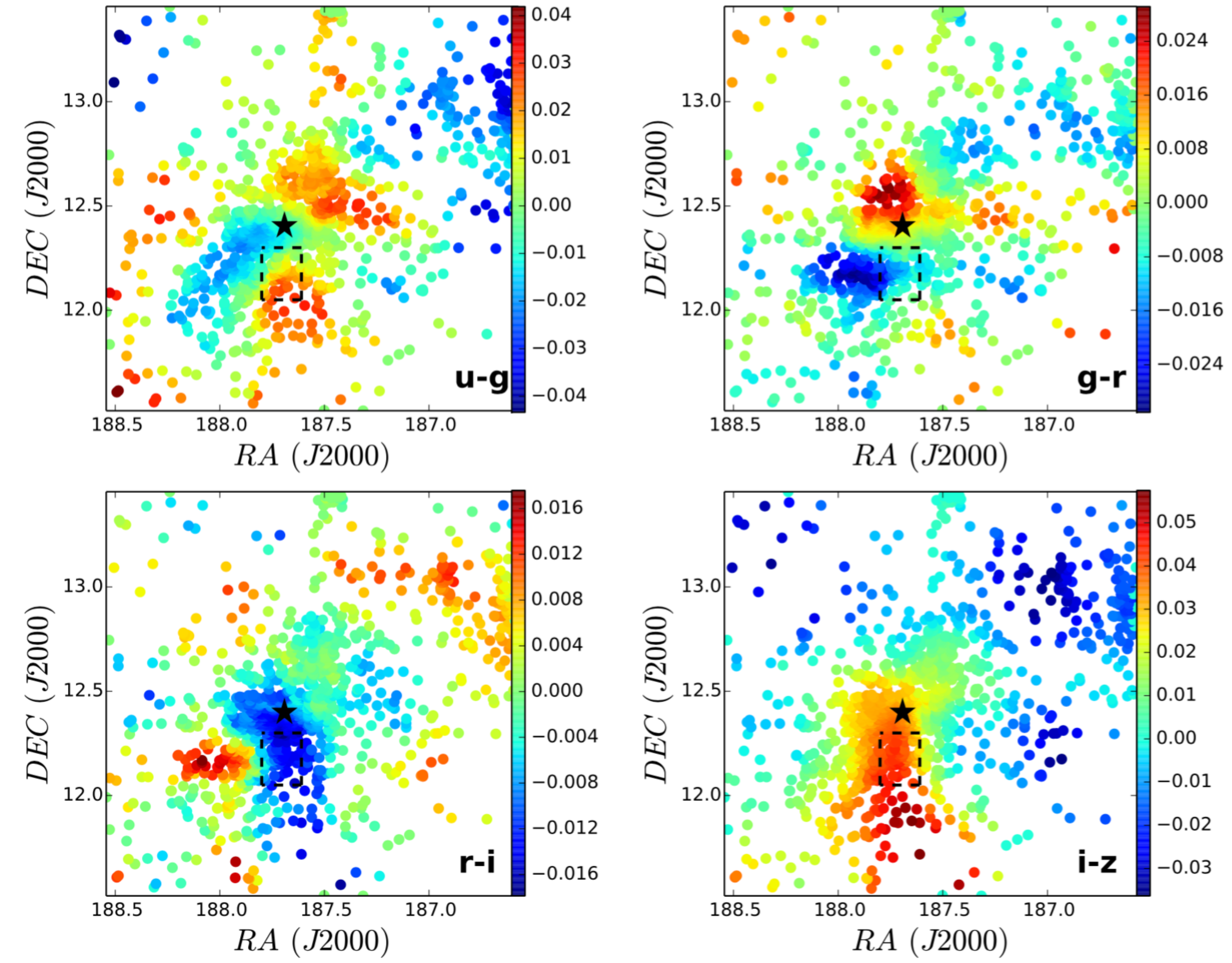}
\caption{GC spatial distribution, color-coded by the stellar color homogeneity correction detailed in Section~\ref{correction}; the symbol colors map the offsets between local and global average stellar color, for the color index given 
at the bottom right corner of each panel.~The black dashed rectangle highlights the position of the potential GC sub-structure discussed in Sections \ref{results} and later.~The black star shows the location of M87, close to the center of the Virgo cluster. \label{fig1}}
\end{figure*}

\begin{figure*}[t]
\begin{center}
\includegraphics[width=\linewidth]{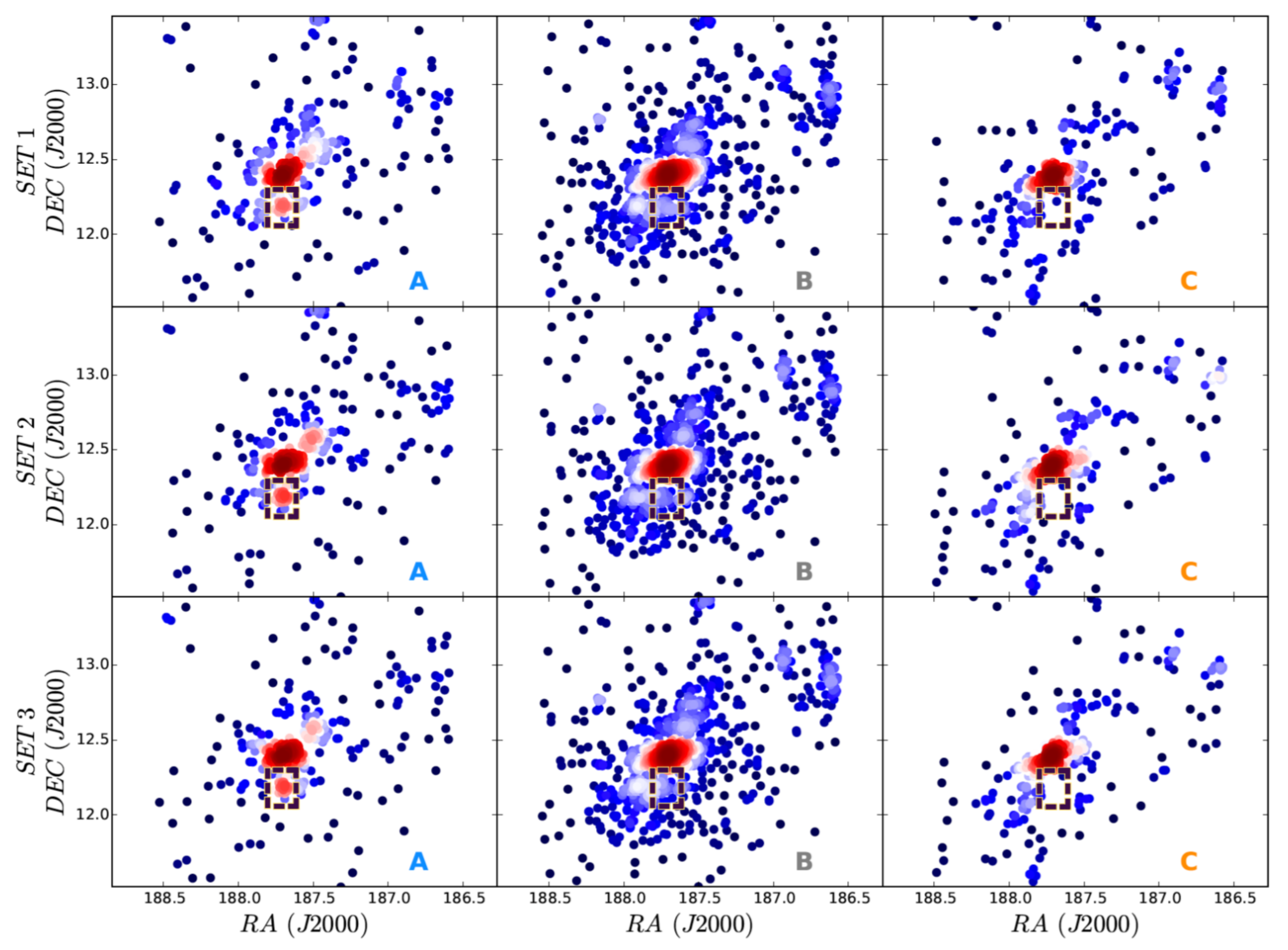}
\caption{Spatial distributions of our sample GCs after separation into three relative age groups (columns for groups $\mathcal{A, B}$ and $\mathcal{C}$, with $\mathcal{A}$ containing the 20\% formally youngest GCs, $\mathcal{C}$ the 20\% formally oldest ones, and $\mathcal{B}$ the 60\% remaining GCs), while each row of the figure is associated with a different set of SSP models.~{\sc Set1} contains three SSP models based on the MILES library, {\sc set2} seven models based on the BaSeL, STELIB, and ATLAS libraries, and {\sc set3} groups all ten models presented in Table\,\ref{tab:model} (see Sect.~\ref{set} for details).~We recall that the formal ages, derived from comparisons between observed colors and SSP models, are not used for their absolute value but only as a particular way of summarizing the position of GCs relative to the typical model locations in color-space.~The display color is related to the local density of GCs, as estimated with a kernel density estimator (the same kernel is used in all panels).~Blue colors indicate low density, while red colors mark high density regions.~The black dashed rectangle south of M87 shows a region containing a relatively dense accumulation of younger GCs.~We also point out the central disky alignment of GC belonging to group $\mathcal{C}$ that are symmetrically distributed around the M87 core.~This structure is aligned with the large-scale GC distribution.~\label{fig2}}
\end{center}
\end{figure*}

\begin{figure*}[t]
\begin{center}
\includegraphics[width=\linewidth]{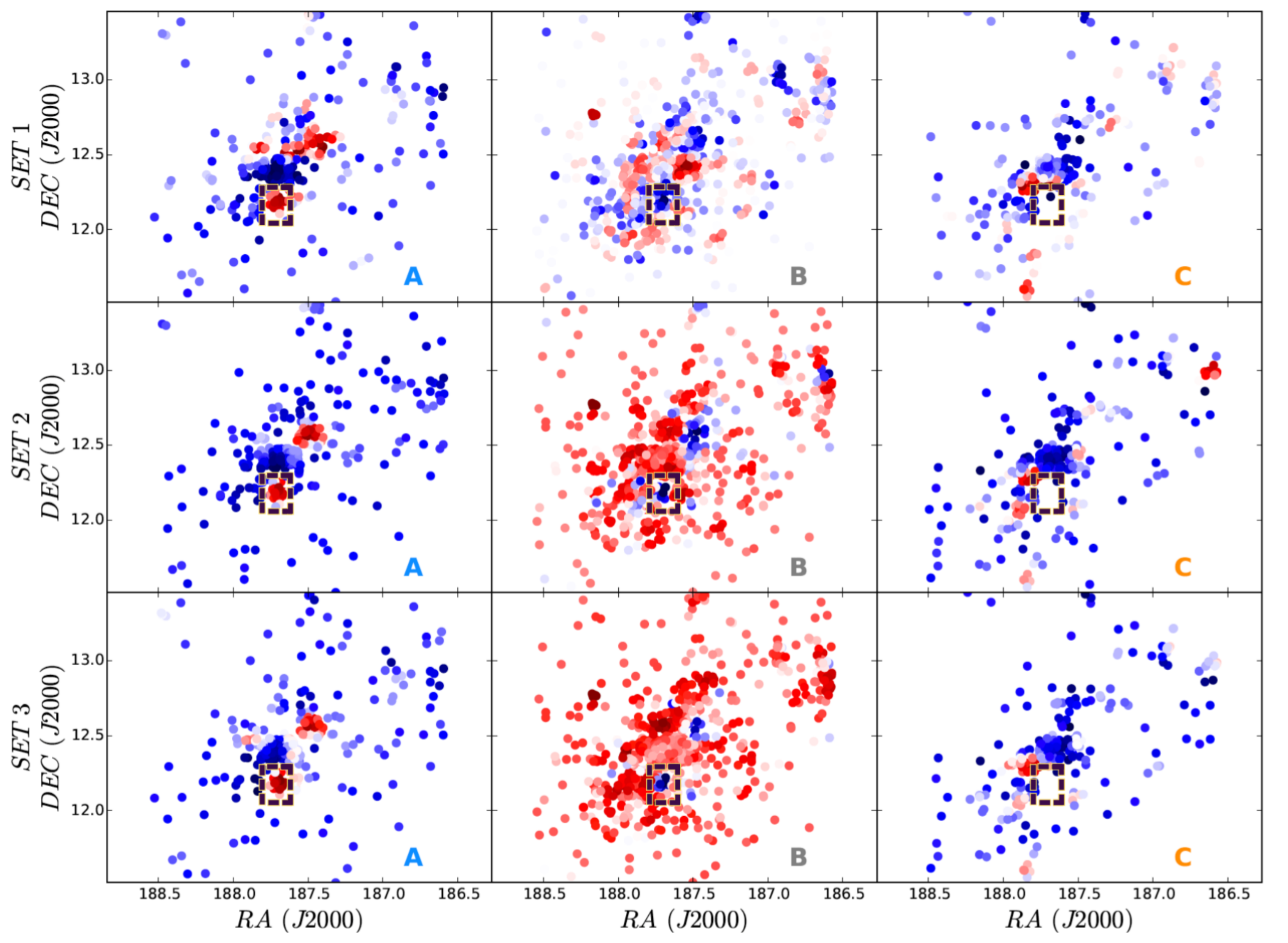}
\caption{Same GC spatial distributions as in Figure~\ref{fig2}, except that this time the color of the symbols encodes the local fraction of the number of GCs of one group ($\mathcal{A, B}$ or $\mathcal{C}$) to the total number of GCs ($\mathcal{A+B+C}$). Red areas represent high fractions, blue areas low fractions. \label{fig3}}
\end{center}
\end{figure*}


\begin{figure}[t]
\centering
\includegraphics[width=\columnwidth]{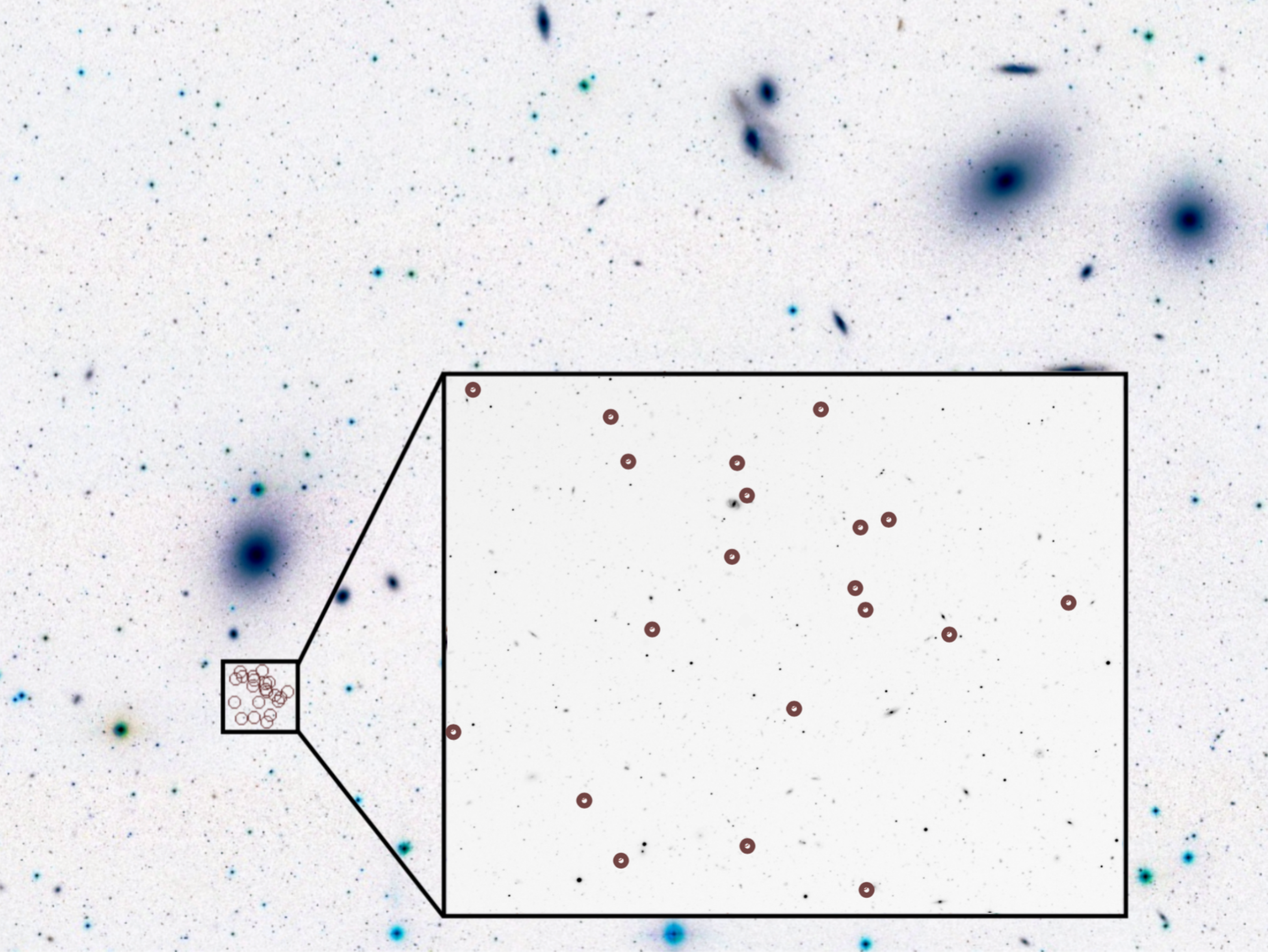}
\caption{Spatial distribution of the potential substructure GCS. The PSGCs (brown circles) are displayed on an NGVS image in the zoom-in panel.~The background image is taken from the Sloan Digital Sky Survey (SDSS).~The three large elliptical galaxies seen in this image are, from the left to the right (i.e. from East to West), M87, M86, and M84.~The zoomed inset covers an area of $6.7\arcmin \times 5.1\arcmin$ (corresponding to $32.2\times24.5$ kpc$^2$).
\label{fig5}}
\end{figure}

\begin{figure*}[t]
\begin{center}
\centerline{\includegraphics[width=0.925\linewidth, trim=0mm 0mm 0mm 0mm]{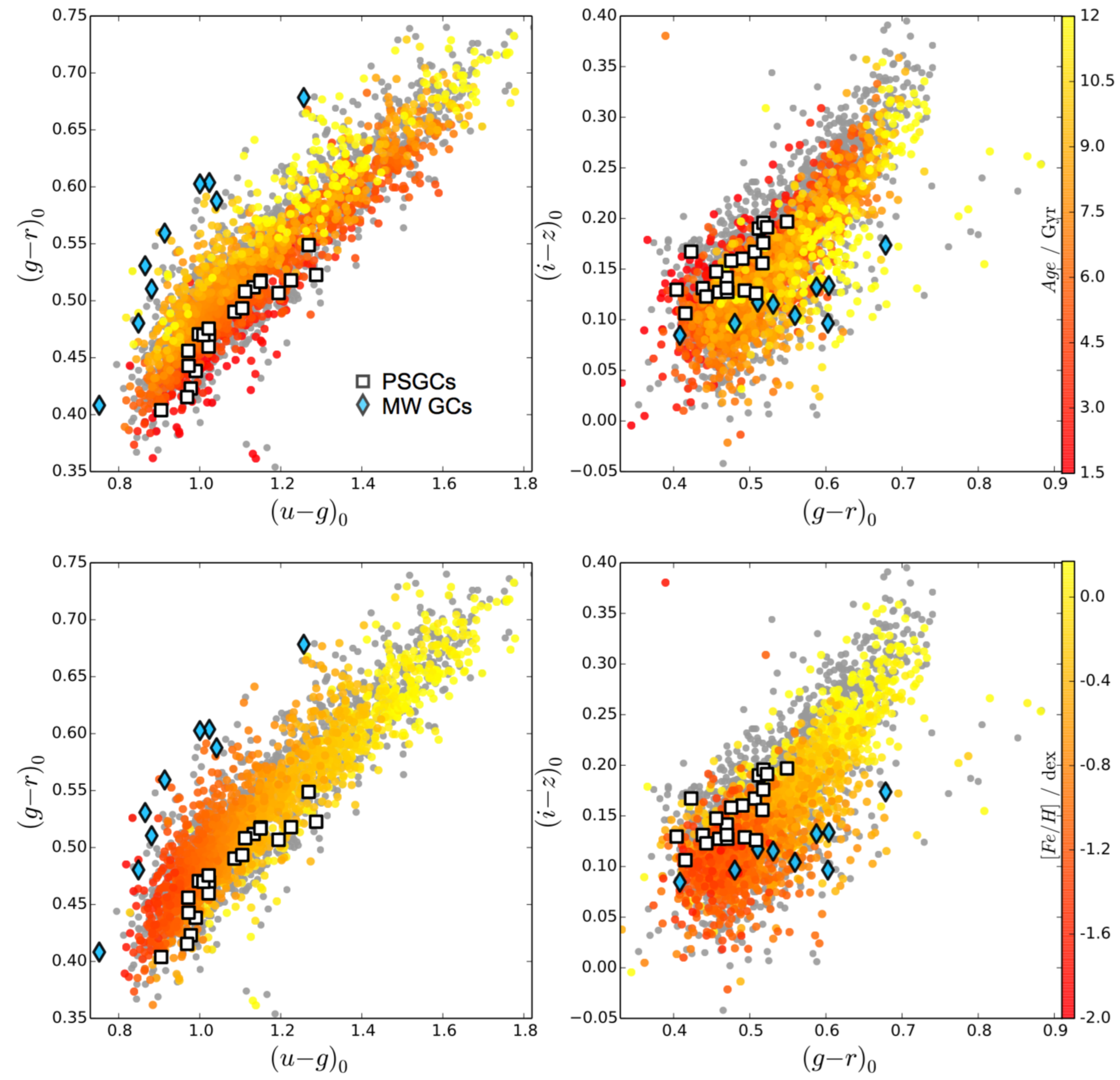}}
\caption{Color-color diagrams for the full NGVS GC sample used in this study. {\it (Top panels)}: $(u\!-\!g)_0$ vs.~$(g\!-\!r)_0$ ({\it left}) and $(g\!-\!r)_0$ vs.~$(i\!-\!z)_0$ color-color diagram ({\it right}), where the symbol colors encode the concordance estimates of ages (see color scale on the right).~The white squares mark the PSGC members of group $\mathcal{A}$, located in the south-east of M87 and discussed in Sect.~\ref{results}. The blue diamonds refer to the Milky Way GCs presented in \citet{powalka2016a}. Finally the grey points show the colors of the NGVS GCs before the correction of Section\,\ref{correction}.~{\it (Bottom panel)}: Similar diagrams as in the top panels, but this time the symbol color is parameterized by the concordance estimates of [Fe/H]~(see color scale on the right). 
\label{fig4}}
\end{center}
\end{figure*}

\section[]{The data: Next Generation Virgo Survey GCs}
\label{sec_data}

\subsection{Colors and magnitudes}
In this paper, we use the high-quality aperture-corrected photometry of the Virgo GC sample defined in Paper~I\footnote{ \scriptsize{\tt{http://vizier.u-strasbg.fr/viz-bin/VizieR?-source=J/ ApJS/227/12 }}}.~This dataset consists of 1846 GCs located within a 3.62~$\rm{deg}^2$ field of view around M87 with photometry in $u^*$, $g$, $r$, $i$, $z$ from CFHT/MegaCam \citep{boulade2003}\footnote{We will not use the $K_s$ photometry here because of its larger random photometric errors.}.~The SExtractor\footnote{\citet{bertin1996}} magnitude random errors in this sample are smaller than 0.06 mag in each band and a budget of systematic errors including errors on calibration, extinction, and filter transmission is presented in Paper~I. The GC sample contains predominantly bright Virgo GCs with mean magnitudes of approximately 23.05, 21.88, 21.32, 21.05, 20.87 in the $u^*$, $g$, $r$, $i$, $z$ filters, respectively.~These magnitudes correspond to typical GC masses of about $2\!\times\!10^6\,M_{\odot}$ at the distance of M87.

\subsection{Relative ages and metallicities}
\label{set}
In Paper~II, we focused on estimating photometric ages and metallicities for the selected Virgo GCs.~We conducted that study by comparing theoretical colors from ten SSP models available in the literature with our observed GC colors.~For any given set of SSP model grids (the ten just mentioned or subsets thereof), we designed a ``concordance estimate" (CE) to assign formal age and metallicity labels to GCs. Those labels extend from 1 Gyr to 14 Gyr for the ages and from $-2.0$ to $0.17$ for the metallicity ([Fe/H]). The CE is defined as the photometric age and metallicity on which those models tend to agree when each of them is used in a simple maximum likelihood analysis (see Paper~II for the precise definition of the adopted robust weighting scheme).~In practice, the resulting values are sensitive to several systematic effects.~The approximation of a GC with a SSP with solar-scaled metal abundances, the model-dependence of predicted colors, and the age-metallicity degeneracy, are amongst the main issues known to affect the final estimates.~Paper~II emphasized how delicate it is to  estimate an {\it absolute} value of age and metallicity accurately.~Therefore, in this letter, we restrict ourselves to the usage of the {\it relative} age and metallicity scale, which is a surrogate of the relative position of each GC in a multi-dimensional color space.

\begin{deluxetable}{lcr}
\tablecaption{\label{tab:model} Stellar libraries and isochrone references for the different SSP models used in this paper. }
\tablehead{
\colhead{Model} 	& \colhead{Stellar library**} 	& \colhead{Isochrones} 	
}
\startdata
BC03 & STELIB & Padova 1994 \\
BC03B & BaSeL 3.1 & Padova 1994   \\
C09BB & BaSeL 3.1 & BaSTI \\
C09PB & BaSeL 3.1 & Padova 2007 \\
C09PM & MILES & Padova 2007 \\
M05 & BaSeL 3.1 & Cassisi \\
MS11 & MILES & Cassisi \\
PEG & BaSeL 2.2 & Padova 1994 \\
PAD & ATLAS ODFNEW / PHOENIX BT-Settl & PARSEC 1.2S  \\
VM12 & MILES* & Padova 2000 \\ \vspace{-3mm}
\enddata
\tablecomments{{\it Models:} BC03 \& BC03B - \citet{bc03}. C09BB, C09PB \& C09PM - (FSPS v2.6) \citet{conroy2009}. M05 \& MS11 - \citet{maraston2005,maraston2011}. PEG - \citet{fioc1997}. PAD - CMD v2.8\footnote{\tt{http://stev.oapd.inaf.it/cgi-bin/cmd}}. VM12 - \citet{vazdekis2012, ricciardelli2012}. {\it Libraries \& isochrones:} ATLAS ODFNEW refers to \citet{castelli2004}. BaSeL: \citet{lejeune1997, lejeune1998, wes02}. BaSTI: \citet{pietrinferni2004,cordier2007}. Cassisi: \citet{cassisi1997a, cassisi1997b, cassisi2000}. MILES: \citet{sanchez2006}. Padova 1994: \citet{alongi1993, bressan1993, fagotto1994a, fagotto1994b, girardi1996}. Padova 2007: \citet{girardi2000, marigo2007, marigo2008}. PARSEC 1.2S: \citet{bressan2012, tang2014, chen2014, chen2015}. PHOENIX BT-Settl: \citet{allard2003}. STELIB: \citet{leb03}.
(*) The MILES library extends from 3464 to 7500\,\AA. In order to reach $u^*$, $i$ and $z$ magnitudes, we used the combinaison of NGSL and MIUSCAT from \citet[][ they provide wavelengths from $\sim$1700\,\AA\ to $\sim$9500\,\AA]{koleva2012} and we extrapolated the NGSL+MILES+MIUSCAT spectra from 9500\,\AA\ to 10000\,\AA\ using Pegase GC spectra.
(**) Optical-only libraries are generally combined with BaSeL at UV and IR wavelengths in the original codes.}
\end{deluxetable}

Following the analysis in Paper~II, we derive for each GC three different CEs of the age and metallicity labels, based on three different subsets of SSP models that are designed to let us search for any dependence of our analysis on the models' input stellar spectral libraries.~We recall the full list of models in Table\,\ref{tab:model}.~The first set ({\sc set1}) contains three SSP models based on the MILES library for the optical wavelengths (C09PM, MS11, VM12).~The second one ({\sc set2}) includes seven models based on the BaSeL, STELIB, and ATLAS libraries (BC03, BC03B, C09BB, C09PB, M05, PAD, PEG).~Finally, the third one ({\sc set3}) groups all ten models.~We use these three sets to search for inhomogeneities in the spatial distribution of the GC spectral energy distributions, which could be signatures of events in the galaxy's past.

\subsection{Stellar color homogeneity}
\label{correction}

The photometric calibration of the NGVS data relies on point sources common to NGVS and to the Sloan Digital Sky Survey (SDSS).~The magnitude range of overlap between these two surveys limits the number of such sources, and hence also the spatial scale on which local deviations between the surveys may be evaluated.~To assess the internal homogeneity of the photometry more precisely, we investigate the distribution of stellar colors within the NGVS pilot field using a larger and deeper stellar sample, this time without the restriction of requiring good SDSS photometry.~We select the stars for this procedure using the $uiK_s$ color-color diagram \citep{munoz2014} together with a measure of source compactness (the latter helps excluding Virgo GCs, as described in Paper~I). 

The top-left panel of Figure~\ref{fig:starselection} illustrates the color magnitude diagram (CMD) of the initial stellar sample, restricted to $18\!<\!g\!<\!25$\,mag to avoid saturation and to limit photometric errors.~Because the survey depth is not uniform in all the photometric bands that define the completeness of this initial catalog, we further restrict the sample everywhere to $20\!<\!g\!<\!22.8$ mag.~This conveniently removes the magnitude range in which the separation between stars and GCs remains difficult even with the $uiK_s$-based method ($g\!>\!23$, and $0.5\!<\!(g\!-\!i)_0\!<\!0.9$).~The cut at $g=22.8$ also avoids a magnitude range where the peak of the main sequence color-distribution moves to redder values.~The properties of the subset are shown in the remaining panels of Figure\,\ref{fig:starselection}.~We then produce RA-DEC maps of the difference, $\Delta$, between the local average color and the global average.~To limit the errors on the estimated local mean, we restrict our sample again, focusing on the peak of the color distribution: $(g\!-\!i)_0\!<\!0.5$\,mag. This cut is wide enough to avoid edge effects, since errors on individual star colors are of a few percent at most, and the local systematics (as we shall see) as well. Changes of $\pm 0.1$ in the color-cut do not modify the results, but the significance with which departures of $\Delta$ from 0 can be estimated drops if the color-range kept is much larger. The local average is estimated within a circle of 0.2\,deg radius, which ensures each estimate is based on 80 to 150 stars except in field corners. The RA-DEC maps of the error on the local mean, $\sigma_{\Delta}$ are mostly flat, with typical values summarized in Table\,\ref{tab:eom}.

\begin{deluxetable}{cccccc}
\tablecaption{\label{tab:eom} Error on the local mean color, in magnitudes, for the star subset defined in the text. }
\tablehead{
\colhead{~} 	& \colhead{$(u-g)_0$} 	& \colhead{$(g-r)_0$} & \colhead{$(r-i)_0$} & \colhead{$(g-i)_0$} & \colhead{$(i-z)_0$} 	
}
\startdata
$\sigma_{\Delta}$ & 0.012 & 0.007 & 0.007 & 0.008 & 0.006 
\enddata
\tablecomments{These values may be exceeded along the edges of the field.}
\end{deluxetable}

Figure~\ref{fig1} shows the local color deviations, $\Delta$, at the locations of globular clusters.
The highest correction values are observed in the $(u\!-\!g)$ and $(i\!-\!z)$ maps with amplitudes between $-0.04$ and $0.06$ mag.~The other colors, $(g\!-\!r)$ and $(r\!-\!i)$, are less affected by the correction, with amplitudes between $-0.02$ and $0.02$ mag.~We observe that the patterns in the $(u\!-\!g)$, $(g\!-\!r)$, and $(i\!-\!z)$ correction maps are independent.~There is some level of anti-correlation between the patterns in $(g\!-\!r)$ and $(r\!-\!i)$, but the two maps combined provide a map of $(g\!-\!i)$ deviations that is not flat and that has a morphology more similar to the $(g\!-\!r)$-map than to the $(r\!-\!i)$-map.

The departures of $\Delta$ from 0 are significant at the $2\sigma_{\Delta}$ level over 25 to 40\!\% of the field, except in $(i\!-\!z)_0$ where the significant deviations cover about 80\!\% of the observed area.~This level of significance was assessed in two ways. First we computed the ratio of $\Delta$ to the uncertainties of that estimated local mean $\sigma_{\Delta}$ (standard deviation in the color / root of the local sample size).~Second, we applied a Kolmogorov-Smirnov (KS) test to local color distributions.~With our selection cuts, we found excellent agreement between regions where $\Delta$ differs from 0 to more than 2\,$\sigma_{\Delta}$, and regions where the assumption of identical global and local color distributions can be rejected at the 95\,\% confidence level via the KS-test.

The causes of the local deviations from the global color distributions can be several.~Subsequently, we assume that zero point errors are dominant.~Hence, we can use the $\Delta$-maps (see Fig.~\ref{fig1}) as a correction to enforce uniform photometry. We will return to the discussion of this assumption in Section\,\ref{discussion}.

In the following, the relative GC ages are estimated after the inclusion of the above correction in the four colors used, i.e. $(u-g)_0$, $(g-r)_0$, $(r-i)_0$ and $(i-z)_0$.

\section{Results}
\label{results}

For each of the model sets, as defined in Section~\ref{set} ({\sc set1} to {\sc set3}), we divide our GC sample into three groups based on the relative formal ages assigned by our CE labelling method.~The first one ($\mathcal{A}$) comprises the 20\% of GCs with the youngest age labels.~The second group ($\mathcal{B}$) contains the intermediate-age GCs (60\% of the GC sample for each set) and the last one ($\mathcal{C}$) takes the 20\% of GCs ranked oldest.

Figure~\ref{fig2} illustrates the spatial distribution of each GC sample $\mathcal{A, B}$ and $\mathcal{C}$, as obtained with each of the three SSP model sets.~The color-code maps the local density of GCs, derived with a gaussian kernel density estimator with the same bandwidth in each of the nine panels.

We notice that the distributions from the {\sc set1, set2}, or {\sc set3} are highly consistent with each other for each age group $\mathcal{A, B}$, and $\mathcal{C}$.~Although each set gives slightly different {\it absolute} ages (see also Paper~II), we observe the same {\it relative} features independently of the model set.~This emphasizes that what we call the 'relative age label' only probes the relative position of any GC in the 4-D color-space, which does not strongly depend on the set of models.~Not surprisingly, we find in each panel an over-density of GCs concentrated around M87, a small region which contains about 50\% of our GC sample.~It is clearly visible at the center of the distribution in each panel.~As the GC distribution is not uniform but exhibits a centrally concentrated number density profile, we additionally present in Figure~\ref{fig3} an alternative color-coding defined by the ratio of the sub-sample density ($\mathcal{A, B}$, or $\mathcal{C}$) by the full sample density ($\mathcal{A+B+C}$).~This plot represents now the proportion of GCs in each of the $\mathcal{A, B}$, or $\mathcal{C}$ samples relative to the full distribution, and can be used to identify candidate regions with an excess of clusters of one or the other category.

Looking at sample $\mathcal{A}$ in Figures~\ref{fig2} and \ref{fig3}, we notice an overdensity of GCs belonging to group $\mathcal{A}$ in the South of M87, centered at ${\rm RA}\!=\!187.7$ and ${\rm DEC}\!=\!+12.2$ and highlighted by a black dashed rectangle.~The absolute difference between the mean CE-age at this location and the mean in the surroundings, equals 5 to 7 times the standard deviation among the surrounding CE-ages (depending on the set of models used), which makes it highly significant.~When counting GCs in ellipses centered on M87, with radially dependent ellipticities and position angles in agreement with values from \citet{durrell2014} and \citet{janowiecki2010}, we find that a randomly selected subset of 20\,\% of the GCs, at the location of the feature of interest and in an area of the same size (i.e. a circular patch of 0.1\,deg diameter), should contain $\Lambda=5^{+2}_{-1}$ objects on average\footnote{To avoid under-estimating the radial number density profile, we re-inject star clusters with poorer $r$-band photometry into the sample, that are present at detector-chip boundaries in one of the quadrants of the Virgo core observations as described in Paper I.}.~
The area of the feature we are observing contains approximately $20$ GCs\footnote{There is a small systematic uncertainty of about $\pm1$ GC, which is due to variations resulting from 1) the particular set of SPS models to compute the ages, and 2) the correction maps using the stellar colors.} of sample $\mathcal{A}$, which is again a significant excess (more than 6\,$\sigma$ when using Poisson statistics for $\Lambda=5$, and still more than 4\,$\sigma$ if $\Lambda=7$).~Surprisingly, as shown in Figure\,\ref{fig5}, we do not find any large Virgo galaxy in this area, which could potentially host this accumulation of GCs.~In the left and middle panels of Figure~\ref{fig3}, we confirm that the area is mainly composed of GCs which are members of group $\mathcal{A}$, unlike its surroundings which are principally occupied by GCs belonging to group $\mathcal{B}$.~These observations hint at a possible GC substructure, with a 4-D position in color space different from those of the neighboring GCs in the (projected) spatial distribution.

Another similar feature is seen to the north-west of M87, around ${\rm RA}\!=\! 187.45$ and ${\rm DEC}\!=\! 12.55$. It has a lower contrast in Figure~\ref{fig2} (left-hand panels) than the Southern feature, but is conspicuous in Figure~\ref{fig3}.~Because it is located in the region of overlap of the corners of the four individual fields of view that were combined to cover the area of our study \citep[see e.g.~Fig.~4 in][]{ferrarese2012}, its photometry is more uncertain even after our corrections.~As a result, the significance of the difference in CE ages between this particular region and its surroundings is low.~Finally, in the right-hand panels of Figure~\ref{fig3}, we notice a few patches with enhanced proportions of GCs of group $\mathcal{C}$.~These regions are difficult to study further because they represent very small numbers of GCs.~Some are associated with galaxies, such as M86 or NGC\,4438 and hence are not the type of features we are  for.

In this letter, we focus on the overdensity to the south of M87 since it has the strongest level of significance.~Within this region, $\sim\!20$ GCs belong to group $\mathcal{A}$ with photometrically computed stellar masses $(0.4-3)\cdot10^6\,M_\odot$.~Hereafter, these 20 GCs are labelled PSGCs (Potential Substructure GCs).~In Figure~\ref{fig4}, we look at their color-color distributions in the context of the full GC sample.~We color-code the symbols with the age and metallicity results from our CE.~The top and bottom panels are, respectively, color-coded by the CE age and metallicity derived with {\sc set1} models.~In each case, both the $(u\!-\!g)_0$ vs.~$(g\!-\!r)_0$ and the $(g\!-\!r)_0$ vs.~$(i\!-\!z)_0$ color-color diagrams are plotted.~PSGCs are highlighted with white squares and we additionally show a sample of Milky Way GCs \citep[defined in][]{powalka2016a} with blue diamonds.~In both color-color diagrams, we observe that the sequence of the PSGCs is tighter than the full GC sample.~Although the gradients of photometric CE age and metallicity have a weak meaning, the PSGC locus seems to follow a relatively young iso-age region in both color-color diagrams, and to be consistent with a range of relatively low metallicities. Their location supports the fact that the overdensity observed in Figures~\ref{fig2} and \ref{fig3} has its cause in the combination of the spatial clustering and the 4-D color-space age coherence. In \citet{powalka2016a} we had already noted that bright Virgo GCs with colors more similar to those of MW GCs, and correspondingly older age labels according to our CEs, are not spatially concentrated but spread over the whole area of the Virgo core field discussed here.

\section{Discussion}
\label{discussion}

Using CE age-labels to characterize the broad-band energy distributions of GCs, and looking at the spatial distribution of those labels around M87, we found a relatively small area to the south of M87 ($\sim\!0.1$\,deg$^2$), where a concentration of GCs consistently exhibits photometric properties different from the GCs in the surroundings. This is an intriguing result, and we start this discussion by re-examining the significance of this detection.~We then consider the origins such a feature may have, if confirmed by future observations.

\subsection{Significance of the GC substructure detection}
\label{significance}

The possible substructure south of M87 was first detected in the spatial maps of the CE ages obtained with the original photometry of Paper~I, i.e.~before the local photometric corrections of Section~\ref{correction}.~In fact, the analysis of the homogeneity of stellar colors and the resulting local photometric spatial uniformity corrections were part of our attempts to explain anomalies in the spatial distributions of GC colors as data reduction artifacts.~While some low-significance substructure in the GC color-maps was eliminated in this process, the candidate substructure that is the focus of this paper could not be erased.~Our local photometry corrections assume that the spatial patterns in Figure~\ref{fig2} are predominantly due to local zero point errors. Here we examine alternative origins, such as selection effects in the star catalogs, real patterns in the stellar colors, local measurement errors due to the presence of extended galaxies, extinction, or the high sensitivity of CE-ages to small changes in color in specific parts of the SED.

\subsubsection{Selection effects}
The depth of the NGVS\,+\,NGVS-IR surveys is not uniform over the field of the Virgo core \citep[especially in $K_s$, which is used for the separation of stars from GCs;][]{munoz2014}.~By carefully selecting a range of stellar magnitudes and colors when calculating the maps of Section~\ref{correction}, we have limited the risk that the patterns may come from spatially varying selection effects.~Failing to apply clean bright and faint magnitude cuts, or to focus on the peak of the $(g\!-\!i)_0$ color distribution, produces maps for $\Delta$ with spatial patterns that are independent of color (except for $[i\!-\!z]$, which remains a special case), showing that selection effects are dominant.~For local zero point errors, we do not expect spatial correlations between stellar color-correction maps that have no passband in common.~The cuts we have applied mostly remove any such spatial correlation between independent color-corrections, while keeping a sample large enough to avoid strong edge-effects and insufficient statistics.

We have already mentioned that changing the exact position of the color and magnitude cuts does not affect the $\Delta$-maps significantly.~In particular, we found that removing the objects bluer than the typical turn-off, e.g.~with $(g\!-\!i)_0\!<\!0.15$, was not critical.~We also found that the significance of the features in the $\Delta$-maps drops when stars with a much broader range of colors, or alternatively only red stars, are kept in the sample (as expected considering the shape of the color distribution in Figure~\ref{fig:starselection}).~Unfortunately, the significance also drops when the radius in which the local estimates of $\Delta$ are computed is reduced below 0.2\,deg, because of insufficient statistics. This remains a limitation of our corrections, although instrumental causes for variations on such small scales are few (remnant effects of gaps between detector chips and individual fields are the main one we are aware of).

\subsubsection{Real patterns in the stellar colors}
The line of sight towards the core of Virgo crosses the Milky Way halo in the direction of the Virgo Stellar Overdensity (VSO) and of a major stream of the Sagittarius dwarf galaxy \citep[][]{jerjen2013, durrell2014, loc16}.~In the range of $g$-band magnitudes we have selected, bright blue stars are thought to belong mostly to the VSO, while fainter ones belong to the Sagittarius stream.~Because the stream and VSO may have spatial substructure, their relative proportions may vary locally and produce differences in $\Delta$ that should not be artificially removed.~However, the average colors of bright and faint stars in our star sample differ by less than 0.008\,mag (in all colors used) which is negligible for our study (and is not measurable locally).

\subsubsection{Local artifacts due to extended galaxies}
The photometry of stars and GCs is sensitive to local background subtraction, in particular near galaxies.~We assessed this using two NGVS source catalogs, one of which contains measurements made after galaxy subtraction \citep[as in][]{alamo17}.~Indeed, colors may vary by up to 0.02\,mag in regions of $\sim\!0.1$\,deg diameter at the location of galaxies.~However, no such changes are found near the feature we are studying here.

\subsubsection{Extinction}
The dereddening of the GC photometry is based on the extinction map of \citet{schlegel1998} as described in Paper~I.~The average extinction value along the line-of-sight towards M87 is small, i.e.~$\langle E_{(B-V)}\rangle\!=\!0.024$\,mag.~In both the \cite{schlafly2011} and the \cite{schlegel1998} extinction maps, we observe that the NGVS central field is located in a line-of-sight unobscured by dust clouds.~The closest cirrus is located $\sim\!1$\,deg south of M87, whereas the PSGCs are centered about 0.2\,deg off M87 in the same direction.~This result is also confirmed in studies of intra-cluster light in the Virgo core region by \citet{rudick2010}.~However, as a sanity check, we have assessed the effect of artificially strengthening or weakening the extinction along the line-of-sight toward the PSGCs.~We repeated our complete analysis with $3\times$ more and $3\times$ less extinction in these directions.~When multiplying the extinction by a factor 3, we only observe a mean age and metallicity variation of 0.3\,Gyr and 0.018\,dex, respectively.~When dividing by the same factor, these variations reach 0.10\,Gyr and 0.03\,dex.~In both cases, the PSGCs remain different from their neighbor GCs and, thus, none of our conclusions are modified.

\subsubsection{Sensitivity of CE estimates to small errors in colors}
Finally, we recall that the transformation from a set of four colors to an age-label is a highly non-linear process.~We tested that small systematic changes in the particular colors of the PSGCs do not lead to large changes in the the age-labels.~The changes applied were vectors in 4-D color-space whose four coordinates took values $-2$, $-1$, $1$ or $2 \times \sigma_{\Delta}$  (all possible combinations). The feature in relative-age maps is robust with respect to such changes: in all cases, the difference between the average age-label within the feature and the average age-label around is more than 5 times the statistical uncertainty on this difference.

\subsection{An astrophysical origin of the GC substructure?}
The dominant galaxy formation scenarios suggest that M87, like other cD galaxies, has grown through multiple mergers and accretions, and that these events have added numerous pre-existing or newly formed GCs to the progenitor \citep[e.g.][]{cote1998, beasley02, har09, renaud2013}.~In a recent study, \cite{ferrarese2016} estimated that as many as 40\,\% of the current M87 GCs could come from disrupted satellites.~This large number lends support to the idea that, within M87, we should observe GCs of various origins with, potentially, significantly different chemodynamical properties.~It is worth examining whether the GC substructure discovered in this work could belong to the remnant GC system of a disrupted galaxy.

\subsubsection{Previously described candidate substructures in the M87 halo}
Previous searches for structure around M87 (other than deep imaging of the diffuse light) have exploited kinematical properties, although these are expensive to obtain.~\citet{romanowsky12} used radial velocities of about 800 globular clusters and searched for wedge-shaped features in phase space (radial velocity vs.~distance to M87) as a signature of co-accreted GC populations.~They identified one candidate structure, for which they excluded chance detection in a random distribution at the 99\,\% level.~In projection on the sky, the GCs in that structure trace a flattened ring-like feature around M87, a shape consistent with the idea of a tidal-disruption event of an infalling satellite galaxy.~Our analysis method could not have detected a population so broadly spread out. The $(g-i)$ colors and spectroscopic metallicity-indicators reported by \citet{romanowsky12} for the GCs associated with the ``wedge" (when available) suggest a broad range of chemical properties, which does not exclude but also does not provide extra support for the accretion picture.

Another candidate structure was reported by \cite{lon15}, based on the radial velocities and positions of planetary nebulae (PNe) around M87.~Bright PNe are associated with the final stellar evolution stages of intermediate-mass stars and, thus, with relatively young stellar populations (a few to several Gyr).~As with the previous study, \cite{lon15} base their detection on wedge-shaped features in phase space, and identify one candidate feature carried by $\sim\!50$ PNe.~In projection on the sky, the PNe are spread all around M87 and are possibly associated with low surface brightness stellar light, in agreement with a picture in which a disrupted companion wraps around the core of M87. Again, this is not the sort of feature our photometric approach could have detected. Because the spatial distribution and velocities are similar but not identical, it remains unclear whether or not this structure and the candidate GC-structure of \citet{romanowsky12} are related to each other.

\subsubsection{The new candidate GC substructure in the context of previous results}
Could the relatively compact candidate GC substructure we have identified be associated with one of the above? The projected distance of our PSGCs to the core of M87 ($\sim \!15\arcmin$, corresponding to $\sim\!72$\,kpc) is compatible with the distances of the kinematically selected candidate structures, but our group of non-typical GCs lies concentrated south of M87, while the spatial distribution of the objects in the two kinematic samples discussed in the previous section (which have to be taken with care because neither are complete) tend to peak on the other side of the galaxy.~Among the GCs with radial velocities cited by \citet{romanowsky12}, for which measurements were made available by \citet{strader2011}, four are located in the area of our structure and three belong to PSGCs, but the authors associate only one of the latter with their ``wedge".~It has a heliocentric radial velocity of $1390\pm17$ km/s. Among the PNe of \citet{lon15}, one is in our area of interest and it is associated with their candidate kinematic structure (with a probability of 80\%).~Its radial velocity is $1287\pm4$ km/s.~Though these two radial velocities are similar, we note that the candidate structure at least in the PN-based case wraps around the galaxy in such a way that both large and small velocities are associated with it in the extended ``tails" globally located in the South of M87 \citep[see Fig.\,2 in][]{lon15}.~The single object that happens to be located in our region of interest is not representative of that diversity. 

\begin{table*}[ht]
\caption[]{Colors and known radial velocities for the PSGCs}
\label{tab:velocities}
\begin{tabular}{cccccccccccccc}
\hline
  \multicolumn{1}{c}{RA} &
  \multicolumn{1}{c}{DEC} &
  \multicolumn{1}{c}{$g$\ mag} &
  \multicolumn{1}{c}{$(u^*\!-\!g)_o$} &
  \multicolumn{1}{c}{$(g\!-\!r)_o$} &
  \multicolumn{1}{c}{$(r\!-\!i)_o$} &
  \multicolumn{1}{c}{$(i\!-\!z)_o$} &
  \multicolumn{1}{c}{$(i\!-\!K_s)_o$} &
  \multicolumn{1}{c}{($u^*\!-\!g)_o$} &
  \multicolumn{1}{c}{$(g\!-\!r)_o$} &
  \multicolumn{1}{c}{$(r\!-\!i)_o$} &
  \multicolumn{1}{c}{$(i\!-\!z)_o$} &
  \multicolumn{1}{c}{$(i\!-\!K_s)_o$} &
  \multicolumn{1}{c}{VEL} \\
  & & & \multicolumn{5}{c}{\rule[0pt]{2cm}{0.4pt} as in Paper I \rule[0pt]{2cm}{0.4pt} } &
           \multicolumn{5}{c}{\rule[0pt]{1.6cm}{0.4pt} after recalibration \rule[0pt]{1.6cm}{0.4pt} } & km.s$^{-1}$ \\
\hline
  187.710 & 12.210 & 20.99 & 1.033 & 0.454 & 0.165 & 0.184 & -0.101 & 1.023 & 0.470 & 0.178 & 0.142 & -0.095 & 1390$^a$\\
  187.657 & 12.188 & 21.80 & 0.996 & 0.413 & 0.205 & 0.202 & -0.071 & 0.978 & 0.423 & 0.219 & 0.167 & -0.062 & 747$^a$\\
  187.676 & 12.183 & 21.61 & 1.107 & 0.48 & 0.208 & 0.199 & 0.084 & 1.087 & 0.490 & 0.222 & 0.160 & 0.089 & \\
  187.696 & 12.218 & 21.92 & 0.916 & 0.393 & 0.166 & 0.172 & -0.208 & 0.905 & 0.405 & 0.179 & 0.131 & -0.208 & \\
  187.691 & 12.190 & 22.11 & 1.007 & 0.425 & 0.205 & 0.172 & -0.085 & 0.990 & 0.437 & 0.218 & 0.131 & -0.083 & \\
  187.708 & 12.205 & 22.32 & 1.144 & 0.496 & 0.201 & 0.232 & 0.225 & 1.133 & 0.512 & 0.215 & 0.190 & 0.232 & \\
  187.690 & 12.199 & 21.60 & 1.24 & 0.505 & 0.233 & 0.236 & 0.175 & 1.225 & 0.518 & 0.246 & 0.195 & 0.178 & \\
  187.728 & 12.146 & 21.91 & 0.991 & 0.404 & 0.205 & 0.153 & -0.175 & 0.969 & 0.415 & 0.215 & 0.107 & -0.172 & \\
  187.729 & 12.217 & 21.99 & 1.295 & 0.508 & 0.288 & 0.236 & 0.228 & 1.287 & 0.522 & 0.300 & 0.191 & 0.233 & \\
  187.726 & 12.210 & 20.98 & 0.981 & 0.427 & 0.189 & 0.166 & -0.181 & 0.972 & 0.443 & 0.202 & 0.122 & -0.175 & 778$^b$\\ 
  187.754 & 12.167 & 21.65 & 1.151 & 0.479 & 0.228 & 0.185 & 0.069 & 1.139 & 0.498 & 0.235 & 0.141 & 0.074 & \\
  187.708 & 12.148 & 22.30 & 1.049 & 0.451 & 0.179 & 0.174 & -0.004 & 1.021 & 0.459 & 0.190 & 0.127 & -0.003 & \\
  187.751 & 12.221 & 22.33 & 1.502 & 0.581 & 0.321 & 0.246 & 0.395 & 1.500 & 0.597 & 0.331 & 0.203 & 0.400 & \\
  187.689 & 12.141 & 20.97 & 1.023 & 0.461 & 0.174 & 0.172 & -0.098 & 0.997 & 0.470 & 0.187 & 0.127 & -0.092 & 1156$^b$\\ 
  187.685 & 12.200 & 21.96 & 1.21 & 0.494 & 0.234 & 0.206 & 0.108 & 1.194 & 0.506 & 0.247 & 0.166 & 0.112 & 1419$^a$\\
  187.700 & 12.170 & 21.90 & 1.174 & 0.507 & 0.203 & 0.221 & 0.169 & 1.150 & 0.517 & 0.216 & 0.176 & 0.170 & \\
  187.689 & 12.186 & 21.85 & 1.167 & 0.505 & 0.213 & 0.197 & 0.109 & 1.148 & 0.516 & 0.226 & 0.155 & 0.111 & \\
  187.734 & 12.156 & 22.59 & 1.43 & 0.57 & 0.293 & 0.232 & 0.316 & 1.410 & 0.584 & 0.302 & 0.187 & 0.320 & \\
  187.710 & 12.194 & 21.93 & 1.022 & 0.455 & 0.186 & 0.174 & -0.072 & 1.008 & 0.469 & 0.199 & 0.131 & -0.067 & \\
  187.723 & 12.183 & 20.65 & 1.127 & 0.476 & 0.214 & 0.169 & 0.021 & 1.112 & 0.491 & 0.225 & 0.124 & 0.023 & 1550$^b$\\ 
\hline\end{tabular} \\
Notes: $a$ - \citet{strader2011}. These 3 objects are also known as [SRB2011]\,H25523, [SRB2011]\,H23419, [SRB2011]\,H24651. $b$ - Spectra obtained by E. W. P. et al. at the MMT Observatory, Mt Hopkins, Arizona.
\end{table*}

Searching the literature and preliminary NGVS-internal catalogs, we find radial velocities for a total of six PSGCs (Table\,\ref{tab:velocities}).~They are spread between 700 and 1900\,km/s.~The dispersion is clearly too large to correspond to a simple dynamically cold system. Hence we suggest two possibilities.~One is that the feature is incidental.~Arguments provided in Sections~\ref{results} and \ref{significance} have excluded this with better than 99\,\% confidence, but we might still be facing the rare accident.~The other possibility is that at least a fraction of the PSGCs belong to a physical group in the process of disruption, which might display a broad range of radial velocities due to a combination of an elongated 3-D spatial configuration and projection effects.~Considering the spatial appearance of the kinematically cold features found previously among GCs and PNe around M87, the PSGCs could be physically associated with GCs dispersed in a wider area, that small number statistics prevent us from identifying unambiguously in our current photometric sample\footnote{In slightly deeper samples based on the NGVS catalogs, number statistics improve somewhat, but the photometric errors reduce the significance of CE-age differences and contaminants rapidly become more troublesome.}. Such a physical group could have preserved a spatial coherence over a few Gyr, but the GCs themselves could be older. Although the PSGS were selected among those with relatively young age labels (among the globally old age range of globular clusters), we remind the reader that young CE-ages are merely a way of describing a particular position in 4-D color-space, relative to standard population synthesis models which tend to reproduce the colors of Milky Way GCs better than those of M87 GCs \citep[see][for details]{powalka2016a}.~As discussed in Paper~II, colors are also sensitive to other stellar population peculiarities (abundance ratios, horizontal branch morphologies, blue stragglers, etc.), and inadequacies of the standard models in these respects are likely to bias the CE-ages.

All in all, the kinematic data available today does not provide conclusive information.~On one hand, previous literature highlights the presence of candidate kinematic structures, that reach the spatial area we are interested in. On the other hand, the evolutionary scenarios associated with the candidate kinematical structures do not explain why there should be a concentration of globular clusters with a specific type of SED to the south of M87. The total number of existing spectroscopic measurements is insufficient to tell us whether or not any particular velocity is over-represented in the area of interest.

\section{Conclusion}
\label{Conclusion}

In this article, we presented a search for potential GC substructures of the rich GC population within a projected radial distance of 0.8$^{\circ}$ of the Virgo core cD galaxy, M87.~We assessed the spatial distribution of the colors of about $\sim\!1800$ luminous GCs with good multi-band photometric measurements from the NGVS (Paper~I).~In addition to the local aperture-corrections already included in the original source catalog, we used a deeper star catalog to improve the homogeneity of point-source colors across our field of view.

After these corrections, we assigned formal relative age and metallicity labels to each GC, using the CE-method described in Paper~II.~Although the absolute ages are subject to large errors (Paper~II), the relative age-labels are good tracers of the relative positions in 4-D color space.~Therefore, we divided the sample into three relative-age bins: sample ($\mathcal{A}$) comprises the 20\,\% of apparently youngest GCs, while sample ($\mathcal{B}$) contains 60\,\% of GCs with intermediate age-labels, and sample ($\mathcal{C}$) includes the remaining 20\,\% of GCs with the oldest age-tags.~These relative-age bins regroup the GCs with similar properties in color-space.

Looking at the spatial distributions of the GCs in each bin, we observed an overdensity of about $20$ GCs in bin $\mathcal{A}$ with a spatial distribution spanning an angular size of $\sim\!0.1$\,deg ($\sim\!30\,$kpc), located to the south of M87 (RA~$\!=\!187.7$ and DEC~$\!=\!+12.2$).~The detection of this structure is robust to any changes in the photometry we deem consistent with systematic errors on the spatial scales at which we can estimate these (a scale about $4\times$ larger than the structure itself).~Surprisingly, in this area, we found no large Virgo galaxies that may potentially host these GCs.~Therefore, we suggest they (or at least some of them) may have been related to a now disrupted satellite of M87.~Unfortunately, kinematic data available to us is as yet too scarce to analyze this assumption in detail.~Candidate dynamical structures identified with GCs by \citet{romanowsky12} or with PNe by \citet{lon15} extend close to the location of our photometric detection, but they are very spread out in space.~Again, exhaustive kinematic data across the face of M87 would be needed to test any direct correlations between these structures and allow for a better description of their morphology. Improving upon current velocity catalogs is difficult, as the GC magnitudes are pushing the limits of 10m-class telescope sensitivities. ~Unless circumstances have led to a drop in merger rates in recent times, the detection of signatures of accretion events around M87 is a natural expectation from cD galaxy formation scenarios, and it is important to continue to search for them.~Our study highlights the importance of well-calibrated ``flat photometry" for searches of spatial substructures in large-field, multi-wavelength imaging surveys. Even deep surveys are limited by the number densities of stars in the halo, when the latter are used for the calibration.~Future spectroscopic campaigns will be necessary to understand the nature of this intriguing GC overdensity.

\acknowledgments
We thank A.~Romanowsky for helpful exchanges on the radial velocities used in 2012.~This project is supported by FONDECYT Regular Project No.~1161817 and BASAL Center for Astrophysics and Associated Technologies (PFB-06), as well as the ECOS-sud/CONICYT French-Chilean collaboration program via project C15U02, and the Institut National des Science de l'Univers of the Centre National de la Recherche Scientifique (CNRS) of France via the Programme National Cosmologie \& Galaxies (PNCG).~EWP acknowledges support from the National Natural Science Foundation of China through Grant No. 11573002.~K.A.M acknowledges support from FONDECYT Postdoctoral Fellowship Project No. 3150599.~E.T acknowledges the support from the Eberhardt Fellowship awarded by the University of the Pacific.

{\it Facilities:} \facility{CFHT (MegaCam/WIRCAM)}, \facility{VLT:Kueyen (X-shooter)}.

\clearpage

\end{document}